\newcommand{\dia}{\!\!\!\!\!\!\!\not\,\,\,\,}

\newcommand{\be}{\begin{eqnarray}}
\newcommand{\ee}{\end{eqnarray}}

\newcommand{\para}{||}
\newcommand{\ppi}{\mathbf{\Pi}}
\documentclass[twocolumn,aps,showpacs]{revtex4}
\usepackage{graphics}
\begin{document}
\title{Chiral fermion mass and dispersion relations at finite
temperature in the presence of hypermagnetic fields}
\author{Alejandro Ayala$^\dagger$, Adnan Bashir$^\ddagger$, Sarira
Sahu$^\dagger$} 
\affiliation{$^\dagger$Instituto de Ciencias Nucleares, Universidad
Nacional Aut\'onoma de M\'exico, Apartado Postal 70-543, M\'exico
Distrito Federal 04510, M\'exico.\\
$^\ddagger$Instituto de F{\'\i}sica y Matem\'aticas,
Universidad Michoacana de San Nicol\'as de Hidalgo, Apartado Postal
2-82, Morelia, Michoac\'an 58040, M\'exico.}

\begin{abstract} 
We study the modifications to the real part of the thermal
self-energy for chiral fermions in the presence of a constant external
hypermagnetic field. We compute the dispersion relation for fermions
occupying a given Landau level to first order in ${g'}^2$, $g^2$
and $g_\phi^2$ and to all orders in $g'B$, where $g'$ and $g$ are the
U(1)$_Y$ and SU(2)$_L$ couplings of the standard model, respectively,
$g_\phi$ is the fermion Yukawa coupling, and $B$ is the
hypermagnetic field strength. We show that in the limit where the
temperature is large compared to $\sqrt{g'B}$, left- and right-handed
modes acquire finite and different $B$-dependent masses due to the
chiral nature of their coupling with the external field. Given the
current bounds on the strength of primordial magnetic fields, we argue
that  the above is the relevant scenario to study the effects of
magnetic fields on the propagation of fermions prior and during the
electroweak phase transition. 
\end{abstract}


\date{\today}

\maketitle

\section{Introduction}

The study of the origin and properties of large scale magnetic fields
has become a subject of intense research over the last
years~\cite{Reviews}. Experimental bounds on their strength can be set
for astrophysical objects whose local electron density and spatial
structure is known or can be estimated~\cite{Kron,
Beck}. For example, in the case of our galaxy, both quantities are
reasonably well known and the average field strength $B$ has been
measured to be on the order of 3 - 4 $\mu$G. Several other spiral
galaxies in the local group contain magnetic fields of similar
intensities. At larger scales, only model dependent upper limits can
be established. These limits are also in the few $\mu$G range. In the
intracluster medium, recent results have shown the existence of $\mu$G
magnetic fields~\cite{Eilek, Clarke}. For intergalactic large scale
fields (dissociated from any particular galaxy or cluster of
galaxies), an upper bound of $10^{-9}$G has been estimated by taking
reasonable values for the magnetic coherence length~\cite{Kron}.  

The origin of these fields is nowadays unknown but it is widely accepted
that in order to generate them, two important ingredients are needed:
a mechanism for creating the seed fields and a process for amplifying
both their amplitude and their coherence scale~\cite{Reviews,
Giovannini1}. Generation of the seed field may be 
either primordial or associated to the process of structure formation.  
During the evolution of the early universe there are a
number of proposed mechanisms that could possibly generate primordial
magnetic fields.  Among the best suited are first order phase
transitions~\cite{Quash, Baym, Boyan}, which provide favorable
conditions for magnetogenesis such as charge separation, turbulence
and out-of-equilibrium conditions. 

Independently of their origin, the presence of primordial magnetic
fields during the evolution of the early universe could have had
important consequences on some cosmological phenomena. For instance,  
magnetic fields can influence big bang nucleosynthesis, thus
affecting the primordial abundance of light elements and the rate of
expansion of the universe. 

Recall that within the standard model (SM) and prior to the
electroweak phase transition (EWPT), the
only magnetic modes able to propagate for large distances belonged to
the (Abelian) U(1)$_Y$ group and are therefore properly called 
{\em hypermagnetic} fields. 
Non-Abelian magnetic fields develop a magnetic
mass through interactions in the electroweak plasma and are thus 
screened. Consequently, fermions coupled chirally to the magnetic
fields through their weak hypercharge. It has recently been shown that
the chiral nature of this coupling is directly responsible for the
building up of an axial asymmetry during the scattering of fermions
with the bubbles of a first order EWPT~\cite{Ayala1, Ayala2}. Another
interesting question is to what extent the hypermagnetic fields
affected the thermal properties of these chiral fermions prior to the
EWPT. This is the issue we take up in this article. In particular, we
study the modification to the mass and dispersion relation of these
chiral fermions in the presence of a constant hypermagnetic field.

This question has been recently explored in Refs.~\cite{Cannellos},
in the limit where $\sqrt{g'B}$ is large compared to the
temperature, where $g'$ is the U(1)$_Y$ coupling
constant. Nevertheless, for homogeneous fields, under the 
assumption of adiabatic amplification, the current bound $B_0\sim
10^{-9}\ $G set by the COBE measurement of temperature
anisotropies~\cite{Barrow}, gives an upper limit $g'B\leq T^2$, at the
EWPT. However, there is no direct bound on the magnetic energy density
as long as this is much smaller than the radiation energy density, and
this condition is satisfied for $T\gg \sqrt{g'B}$. In any case, the
current observations seem to favor the scenario in which $T$ is larger
--if not much larger-- than $\sqrt{g'B}$, prior to and during the EWPT.

\begin{figure}[t!] 
\vspace{0.5cm}
{\centering
\resizebox*{0.4\textwidth}{0.2\textheight}
{\includegraphics{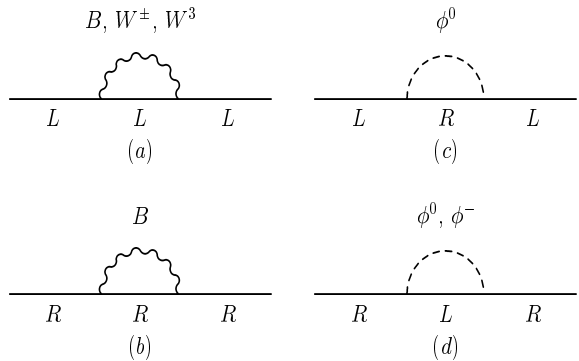}}
\par}
\caption{Feynman diagrams contributing to the chiral fermion
self-energy at one loop.}
\label{fig1}
\end{figure}

In this work we compute the self-energies for left- and
right-handed fermions in the relevant context of the SM for
temperatures prior to the EWPT in the presence of hypermagnetic
fields. We use Schwinger's proper time method~\cite{Schwinger, Sahu} to
incorporate the effects of the external field to all orders in $g'B$
in the propagators and work at one loop level. By considering the
limit $T\gg\sqrt{g'B}$ we find the dispersion relations for these
fermions. We show that for a given Landau level, these modes
acquire finite and different $B$-dependent thermal masses due to the
chiral nature of the fermion coupling to the hypermagnetic field. We
also find the mass splitting between particle and collective (hole)
excitations. In terms of the kinematice regimes of 
temperature and magnetic fields, our work compliments the one in
Ref.~\cite{Cannellos}. 
 
The work is organized as follows: In Sec.~\ref{II} we compute the
finite temperature fermion self-energies at one-loop in the presence
of an external hypermagnetic field using Schwinger's proper time
method. In Sec.~\ref{III}, we work in the weak field limit and in
Sec.~\ref{IV} we further consider the high-temperature limit.
In Sec.~\ref{V} we find the mass and dispersion relations for left-
and right-handed modes. Although the sample calculation presented there 
has been worked out for the top quark alone, the formalism presented is
general and can be readily applied to the case of any SM fermion. 
We summarize and discuss our results in 
Sec.~\ref{concl}. 

\section{Chiral fermion self-energies in a constant hypermagnetic
field}\label{II} 

The Feynman diagrams contributing to the one-loop left- and
right-handed fermion self energies in the symmetric phase of the SM
are depicted in Fig.~\ref{fig1}. Figures~\ref{fig1}($c$) and~($d$)
represent the contributions from internal Higgs boson lines and 
are thus proportional to $g^2_\phi$, where $g_\phi$ is the Yukawa
coupling for the given fermion species. Since the Yukawa couplings
and the vacuum masses of fermions are proportional to each other, the
contribution from these diagrams is only significant for the top quark
for which $g_\phi\sim {\mathcal {O}}(1)$. Since we are interested in
describing the thermal properties of fermions propagating in the early
universe where the particle antiparticle asymmetry is small, we
consider that the chemical potential vanishes and thus that there is
no contribution to the self-energies from tadpole diagrams.

In order to consider the effect of the external field to all
orders in $\sqrt{g'B}$, we need to write the exact propagators in the
presence of a constant hypermagnetic field. Notice that gauge boson
propagators in the internal lines of Fig.~\ref{fig1} $(a)$ and~$(b)$
do not couple to an external U(1)$_Y$ field and thus do not need to be
dressed by the effects of this field. We take ${\mathbf {B}}$
oriented in the $\hat{z}$ direction. Using Schwinger's proper-time
method, it is possible to obtain the exact expressions for the
vacuum propagators of the massless (hyper)charged fermion
and scalar-boson, $S_0(x',x'')$ and $D_0(x',x'')$, respectively
\be
   \!\!\!\!S_0(x',x'')&=&
   \phi (x',x'')\int\frac{d^4k}{(2\pi^4)}e^{-ik\cdot
   (x'-x'')}S_0(k)\nonumber\\
   \!\!\!\!D_0(x',x'')&=&
   \phi^* (x',x'')\int\frac{d^4k}{(2\pi^4)}e^{-ik\cdot
   (x'-x'')}D_0(k),
   \label{props}
\ee
where $S_0(k)$ and $D_0(k)$ are given by
\begin{widetext}
\be
   iS_0(k)&\equiv&
   \int_0^\infty\frac{ds}{\cos z}
   \exp\left[is\left(k_{\para}^2 -k_\perp^2\ 
   \frac{\tan z}{z}\right)-\epsilon |s|\right]
   \left\{k\;\!\dia_{\para} e^{iz\Sigma_3}
   - \frac{k\;\!\dia_\perp}{\cos z}\right\}\nonumber\\
   iD_0(k)&\equiv&
   \int_0^\infty\frac{ds}{\cos\tilde{z}}
   \exp\left[is\left(k_{\para}^2 -k_\perp^2\ 
   \frac{\tan\tilde{z}}{\tilde{z}}\right)
   -\epsilon |s|\right]\, ,
   \label{defsS0D0}
\ee
\end{widetext}
where for simplicity we write $z=g'YBs$ and $\tilde{z}=Y_\phi Bs/2$,
with $Y=Y^{R,L}/2$ the hypercharge for the left- or
right-handed mode of the given fermion species, and $Y_\phi$ the
hypercharge for the boson field. Also, in Eq.~(\ref{defsS0D0}) we use
the definitions
\be
   (a\cdot b)_{\para}&=&a^0b^0-a^3b^3\nonumber\\
   (a\cdot b)_{\perp}&=&a^1b^1+a^2b^2\, ,
\ee
and
\be
   \Sigma_3\equiv\sigma^{12}=\frac{i}{2}[\gamma^1,\gamma^2]
   =\left(
   \begin{array}{cc}
      \sigma_3&0\\
      0&\sigma_3
   \end{array}\right)\, .
   \label{sig3}
\ee

The vacuum gauge boson propagator is simply given by 
\be
   G_0^{\mu\nu}(q)=-\frac{1}{q^2+i\epsilon}
   \left[g^{\mu\nu}-\left(1-\frac{1}{\xi}\right)\frac{q^\mu q^\nu}{q^2}
   \right]\, ,
   \label{gaugebosprop}
\ee
where $\xi$ is the gauge-fixing parameter. The phase factor $\phi$ in
Eqs.~(\ref{props}) is given by
\be
   \phi (x',x'')=\exp\left[ig'Y\int_{x''}^{x'}dx_\mu A^\mu 
   (x)\right]\, .
   \label{phase}
\ee
and does not depend on the integration path. It depends however on the
choice of the residual gauge freedom. We work in the {\it symmetric}
gauge $A_\mu=-\frac{\mbox\small{1}}{\mbox\small{2}}F_{\mu\nu}x^\nu$
and thus
\be
   \phi (x',x'')=\exp\left(i\frac{g'Y}{2}{x_\mu}''F^{\mu\nu}{x'}_\nu
   \right)\, .
   \label{phaseexp}
\ee
We use the real-time formulation of thermal field theory, writing the
fermion, scalar-boson and gauge-boson propagators at finite 
temperature as
\be
   S(x',x'')&=&\phi (x',x'')\int\frac{d^4k}{(2\pi^4)}e^{-ik\cdot
   (x'-x'')}S(k)\nonumber\\
   D(x',x'')&=&\phi^* (x',x'')\int\frac{d^4k}{(2\pi^4)}e^{-ik\cdot
   (x'-x'')}D(k)\nonumber\\
   G(x',x'')&=&\int\frac{d^4k}{(2\pi^4)}e^{-ik\cdot
   (x'-x'')}G(k)\, ,
   \label{propsfinT}
\ee
where 
\be
   S(k)&=&S_0(k)-\tilde{n}_F(k_0)[S_0(k)-\bar{S}_0(k)]\nonumber\\
   D(k)&=&D_0(k)+n_B(k_0)[D_0(k)-D_0^*(k)]\nonumber\\
   G^{\mu\nu}(k)&=&G^{\mu\nu}_0(k)+n_B(k_0)
   [G^{\mu\nu}_0(k)-G^{\mu\nu*}_0(k)]\, .
   \label{propsrealtime}
\ee
In Equations~(\ref{propsrealtime}),
$\tilde{n}_F(k_0)=\tilde{n}(k_0)\theta (k_0) + \tilde{n}(-k_0)\theta
(-k_0)$ and $n_B=n(k_0)\theta (k_0) + n(-k_0)\theta (-k_0)$ where 
\be
   \tilde{n}(k_0)&=&\frac{1}{e^{k_0/T}+1}\nonumber\\
   n(k_0)&=&\frac{1}{e^{k_0/T}-1}\, 
   \label{distribs}
\ee
are the Fermi-Dirac and Bose-Einstein distributions, respectively.
In analogy to Eqs.~(\ref{propsfinT}), the corresponding expression for
the fermion self-energy is given by
\be
   \Sigma(x',x'')=\phi (x',x'')\int\frac{d^4k}{(2\pi^4)}e^{-ik\cdot
   (x'-x'')}\Sigma(k)\, .
   \label{sigmaspace}
\ee
The contribution to the fermion self-energy stemming from the exchange
of gauge bosons, depicted in Figs.~\ref{fig1}($a$) and~($b$) can be
generically written as  
\begin{widetext}
\be
   -i\Sigma^{\mbox{\tiny{gauge}}}(k)=iG^2
   \int\frac{d^4p}{(2\pi)^4}
   \left\{
   \begin{array}{c}
      R \\ 1
   \end{array}
   \right\}
   \gamma_\mu S(p)G^{\mu\nu}(q)\gamma_\nu
   \left\{
   \begin{array}{c}
      L \\ 1
   \end{array}
   \right\}
   \, ,
   \label{siggauge}
\ee
\end{widetext}
where $R=\frac{\mbox\small {1}}{\mbox\small {2}}(1+\gamma_5)$,
$L=\frac{\mbox\small {1}}{\mbox\small {2}}(1-\gamma_5)$ and $G$
stands for $\frac{\mbox\small {g}}{\mbox\small {2}}$ or
$g'Y$. The upper row in Eq.~(\ref{siggauge}) refers to 
the case where the external fermion is left-handed and the lower row
corresponds to the case where the external fermion is right-handed. On
the other hand, the contribution to the fermion self-energy coming
from the exchange of a scalar boson, depicted in Figs.~\ref{fig1}
$(c)$ and~$(d)$ can be generically written as 
\be
   -i\Sigma^{\mbox{\tiny{scalar}}}(k)=g_\phi^2
   \int\frac{d^4p}{(2\pi)^4}S(p)D(q)\, ,
   \label{sigscalar}
\ee
In Eqs.~(\ref{siggauge}) and~(\ref{sigscalar}), $q=k-p$. 

Since we are interested in computing the dispersion relation, we look
at the real parts of Eqs.~(\ref{siggauge}) and~(\ref{sigscalar}) and
thus 
\be
   &&i\int\frac{d^4p}{(2\pi)^4}\gamma_\mu S(p)G^{\mu\nu}(q)\gamma_\nu
   \rightarrow\Sigma_W(k) + \Sigma_f(k)\nonumber\\
   &&i\int\frac{d^4p}{(2\pi)^4}
   S(p)D(q)\rightarrow\Sigma_\phi(k) + \Sigma_{f\phi}(k)\, ,
   \label{realparts}
\ee
where we have defined
\be
   \Sigma_W(k)\!\!&\equiv&\!\!-2\pi\int\!\!\frac{d^4p}{(2\pi)^4}
   \gamma_\mu S_0(p)\gamma^\mu\delta (q^2)n_B(q)\nonumber\\
   \Sigma_f(k)\!\!&\equiv&\!\!i\int\!\!\frac{d^4p}{(2\pi)^4}
   \gamma_\mu\frac{[S_0(p)-\bar{S}_0(p)]}{q^2+i\epsilon}\gamma^\mu
   \tilde{n}_F(q)\nonumber\\
   \Sigma_\phi(k)\!\!&\equiv&\!\!i\int\!\!\frac{d^4p}{(2\pi)^4}
   S_0(p)[D_0(q)-D_0^*(q)]n_B(q)\nonumber\\
   \Sigma_{f\phi}(k)\!\!&\equiv&\!\!-i\int\!\!\frac{d^4p}{(2\pi)^4}
   D_0(q)[S_0(p)-\bar{S}_0(p)]\tilde{n}_F(p)
   \label{defsigs}
\ee
and in the first of Eqs.~(\ref{realparts}) we have left out the
term proportional to $(1-1/\xi)$ since in the large temperature limit
its contribution will be sub-dominant. 

In what follows, we work out explicitly the expression for $\Sigma_W$
and quote the result for the rest of the terms contributing to
Eqs.~(\ref{realparts}). 

We first introduce the representations
\be
   \delta (q^2)&=&\frac{1}{2\pi}
   \int_{-\infty}^\infty dt e^{i(tq^2+i|t|\epsilon )}\nonumber\\
   \frac{1}{q^2+i\epsilon}&=&-i
   \int_{0}^\infty dt e^{it(q^2+i\epsilon )}\, .
   \label{rep}
\ee 
Using the first of Eqs.~(\ref{props}) into the expression for
$\Sigma_W (k)$ and after performing the Gaussian integrations, we
obtain
\begin{widetext}
\be
   \Sigma_W(k)&=&2\int\frac{dp_0}{(2\pi)^4}n_B(q_0)
   \int_{-\infty}^\infty dt\int_0^\infty ds
   \left(\frac{\pi}{i[s+t]}\right)^{1/2}
   \left(\frac{\pi}{i[\frac{s\tan z}{s}+t]}\right)\sec^2 z\nonumber\\
   &&
   \exp\left\{i\left[ sp_0^2 + t(p_0-k_0)^2- 
   \left(\frac{ts}{s+t}\right)k_3^2 
   - \left(\frac{t\tan z}{\tan z +
   YBt}\right)k_\perp^2\right]-(|t|+|s|)\epsilon\right\}\nonumber\\
   &&\left\{\left[p_0u\;\dia + \left(\frac{t}{s+t}\right)
   k_3b\;\dia\right]\exp\left\{ig'YBs\Sigma_3\right\}\cos z
   -\left(\frac{t}{\frac{s\tan z}{z}+t}\right)
   (\gamma\cdot k)_\perp\right\}\, ,
   \label{SigmaWne}
\ee
\end{widetext}
where we have introduced the definitions
\be
   u^\mu&=&(1,0,0,0)\nonumber\\
   b^\mu&=&(0,0,0,1)\, .
   \label{defsint}
\ee
Equation~(\ref{SigmaWne}) is not an operator but rather a complicated
function of $k$. There is a way however to convert it into a gauge
invariant operator~\cite{Schwinger, Dittrich} which, when acting on
the wave functions of interest (see Sec.~\ref{V}) has simple
properties. 

Notice that since $\phi (x',x'')$ depends only on $x_\perp'$ and
$x_\perp''$, we can write
\be
   \hat{\Sigma} (x',x'')&\equiv&\langle x'|\hat\Sigma
   |x''\rangle\nonumber\\
   &=&\int\frac{d^2k_{\para}}{(2\pi)^2}e^{-ik\cdot(x'-x'')_{\para}}
   \nonumber\\
   &&\left\{\!\!\phi (x',x'')\!\!\!\int\!\!\frac{d^2k_\perp}{(2\pi)^2}
   e^{ik\cdot(x'-x'')_\perp}\Sigma (k)\!\!\right\}.
   \label{oper}
\ee
We now use the relations~\cite{Dittrich}
\begin{widetext}
\be
   \phi
   (x',x'')\int\frac{d^2k_{\perp}}{(2\pi)^2}e^{ik\cdot(x'-x'')_{\perp}}
   \exp\left[-i\frac{\tan v}{g'YB}k_\perp^2\right]\!\!&=&\!\!
   \langle x'|\exp\left[-i\frac{v}{g'YB}\ppi_\perp^2\right]|x''\rangle
   \cos v\nonumber\\ 
   \phi
   (x',x'')\int\frac{d^2k_{\perp}}{(2\pi)^2}
   e^{ik\cdot(x'-x'')_{\perp}} 
   \exp\left[-i\frac{\tan v}{g'YB}k_\perp^2\right]
   (\gamma\cdot k)_\perp\!\!&=&\!\!\langle
   x'|\exp\left[-i\frac{v}{g'YB}(\ppi_\perp^2-g'YB\Sigma_3)\right]
   \ppi\dia_\perp |x''\rangle
   \cos^2 v\, ,
   \label{identities}
\ee
\end{widetext}
where $\ppi_\mu = k_\mu - g'Y A_\mu$ and we have introduced the
definition 
\be
   \tan v&=& \frac{g'YBt\tan z}{\tan z + g'YBt}\, .
   \label{deftanv}
\ee
Using Eqs.~(\ref{sigmaspace}),~(\ref{SigmaWne})
and~(\ref{identities}), we obtain
\begin{widetext}
\be
   \hat{\Sigma}_W&=&2\int\frac{dp_0}{(2\pi)^4}n_B(q_0)
   \int_{-\infty}^\infty dt\int_0^\infty ds
   \left(\frac{\pi}{i[s+t]}\right)^{1/2}
   \left(\frac{\pi}{i[\frac{s\tan z}{s}+t]}\right)\sec^2\!z\cos v\nonumber\\
   &&
   \exp\left\{i\left[sp_0^2 + t(p_0-k_0)^2- \left(\frac{ts}{s+t}\right)k_3^2 
   - \left(\frac{v}{g'YB}\right)\ppi_\perp^2\right]-(|t|+|s|)\epsilon\right\}
   \nonumber\\
   &&\left\{\left[p_0u\;\dia + \left(\frac{t}{s+t}\right)
   k_3b\;\dia\right]\exp\left\{ig'YBs\Sigma_3\right\}\cos z
   -\left(\frac{t\cos v}{\frac{s\tan z}{z}+t}\right)
   \exp\left\{i\Sigma_3 v\right\}
   \ppi\dia_\perp\right\}\, .
   \label{SigmaWneop}
\ee
\end{widetext}
The result obtained so far is valid for any strength of the magnetic
field. In order to pursue our objective though, in the next section we
explore the weak field limit.

\section{Weak field limit}\label{III}

In anticipation to considering the limit $T\gg\sqrt{g'B}$, let us
expand the integrand in Eq.~(\ref{SigmaWneop}) up to linear order in
$B$. Care has to be taken when considering the terms where $B$ is
added linearly to $\ppi^2$, since, as we will show below
(Sec.~\ref{V}), for Landau levels with a large principal quantum
number $n$, $\ppi^2\simeq k_3^2+g'Y^{R,L}Bn$. For these terms we thus
keep the full dependence on $B$.

After expanding to linear order in $B$ in the appropriate terms and
performing the shift $k_0-p_0\rightarrow p_0$ in the variable of
integration, we get
\begin{widetext}
\be
   \hat{\Sigma}_W&=&\frac{1}{8\pi^{5/2}}
   \int_{-\infty}^\infty dp_0n_B(p_0)\int_{-\infty}^\infty dt
   \int_0^\infty ds\left(\frac{1}{i[s+t]}\right)^{3/2}
   \exp\left\{ i(tp_0^2+s(k_0-p_0)^2-\frac{st}{s+t}\ppi^2)-(|t|+|s|)
   \epsilon\right\}\nonumber\\
   &&
   \left\{\left[(k_0-p_0)u\;\dia + \left(\frac{t}{s+t}\right)
   k_3b\;\dia\right]\exp\left\{ig'YBs\Sigma_3\right\}
   -\left(\frac{t}{s+t}\right)\exp\left\{i\left(\frac{st}{s+t}
   \right)g'YB\Sigma_3\right\}
   \ppi\dia_\perp\right\}\, ,
   \label{sigmaW1}
\ee
\end{widetext}
where $\ppi^2=\ppi_\perp^2 + k_3^2$.
To carry out the integrations over $s$ and $t$, for the integration
region $0\leq s\leq\infty$, $0\leq t\leq\infty$ we make the change of
variables
\be
   t&=&xy\nonumber\\
   s&=&x(1-y)\, ,
   \label{change1}
\ee
and for the integration region $0\leq s\leq\infty$, $-\infty\leq
t\leq 0$ we make the change
\be
   t&=&xy\nonumber\\
   s&=&x(1+y)\, .
   \label{change2}
\ee
After performing the integrals over $x$ and $y$ we obtain
\begin{widetext}
\be
   \hat{\Sigma}_W&=&\frac{1}{8\pi^2}
   \int_{-\infty}^\infty dp_0 n_B(p_0)\left\{
   \left[\frac{(k_0-p_0)u\;\dia}{\sqrt{\ppi^2}} 
   + \frac{k_3b\;\dia}{2(\ppi^2)^{3/2}}\
   \left(p_0^2-(k_0-p_0)^2-\ppi^2-g'YB\Sigma_3\right)\right]\right.\nonumber\\
   &&\ln\left|\frac{p_0^2-(k_0-p_0)^2-\ppi^2-g'YB\Sigma_3
   -2\sqrt{\ppi^2}\sqrt{(k_0-p_0)^2+g'YB\Sigma_3}}
   {p_0^2-(k_0-p_0)^2-\ppi^2-g'YB\Sigma_3
   +2\sqrt{\ppi^2}\sqrt{(k_0-p_0)^2+g'YB\Sigma_3}}\right|\nonumber\\
   &+&\frac{\ppi\dia_\perp}{2(\ppi^2-g'YB\Sigma_3)^{3/2}}\
   \left(p_0^2-(k_0-p_0)^2-\ppi^2+g'YB\Sigma_3\right)\nonumber\\
   &\times&
   \ln\left|\frac{p_0^2-(k_0-p_0)^2-\ppi^2+g'YB\Sigma_3
   -2\sqrt{\ppi^2}\sqrt{(k_0-p_0)^2+g'YB\Sigma_3}}
   {p_0^2-(k_0-p_0)^2-\ppi^2+g'YB\Sigma_3
   +2\sqrt{\ppi^2}\sqrt{(k_0-p_0)^2+g'YB\Sigma_3}}\right|\nonumber\\
   &+&\left.\frac{2k_3b\;\dia}{\ppi^2}\sqrt{(k_0-p_0)^2+g'YB\Sigma_3}
   +\frac{2\ppi\dia_\perp}{\ppi^2-g'YB\Sigma_3}\sqrt{(k_0-p_0)^2}\right\}
   \, .\label{SigmaW2}
\ee
\end{widetext}
In a similar fashion, we can compute the corresponding expressions for
$\hat{\Sigma}_f$, $\hat{\Sigma}_\phi$ and $\hat{\Sigma}_{f\phi}$ with
the results 
\begin{widetext}
\be
   \hat{\Sigma}_f&=-&\frac{1}{8\pi^2}
   \int_{-\infty}^\infty dp_0 \tilde{n}_F(p_0)\left\{
   \left[\frac{p_0u\;\dia}{\sqrt{\ppi^2}} 
   + \frac{k_3b\;\dia}{2(\ppi^2)^{3/2}}\
   \left(p_0^2-(k_0-p_0)^2+\ppi^2+g'YB\Sigma_3\right)\right]\right.\nonumber\\
   &&\ln\left|\frac{p_0^2-(k_0-p_0)^2-\ppi^2+g'YB\Sigma_3
   -2\sqrt{\ppi^2}\sqrt{(k_0-p_0)^2}}
   {p_0^2-(k_0-p_0)^2-\ppi^2+g'YB\Sigma_3
   +2\sqrt{\ppi^2}\sqrt{(k_0-p_0)^2}}\right|\nonumber\\
   &-&\frac{\ppi\dia_\perp}{2(\ppi^2+g'YB\Sigma_3)^{3/2}}\
   \left(p_0^2-(k_0-p_0)^2+\ppi^2+g'YB\Sigma_3\right)\nonumber\\
   &\times&
   \ln\left|\frac{p_0^2-(k_0-p_0)^2-\ppi^2-g'YB\Sigma_3
   -2\sqrt{(k_0-p_0)^2}\sqrt{\ppi^2+g'YB\Sigma_3}}
   {p_0^2-(k_0-p_0)^2-\ppi^2-g'YB\Sigma_3
   +2\sqrt{(k_0-p_0)^2}\sqrt{\ppi^2+g'YB\Sigma_3}}\right|\nonumber\\
   &-&\left.\frac{2k_3b\;\dia}{\ppi^2}\sqrt{(k_0-p_0)^2}
   -\frac{2\ppi\dia_\perp}{\ppi^2+g'YB\Sigma_3}\sqrt{(k_0-p_0)^2}\right\}
   \, ,\nonumber\\
   \hat{\Sigma}_\phi&=&\frac{1}{16\pi^2}
   \int_{-\infty}^\infty dp_0 n_B(p_0)\left\{
   \left[\frac{(k_0-p_0)u\;\dia}{\sqrt{\ppi^2}} 
   + \frac{k_3b\;\dia}{2(\ppi^2)^{3/2}}\
   \left(p_0^2-(k_0-p_0)^2-\ppi^2+g'YB\Sigma_3\right)\right]\right.\nonumber\\
   &&\ln\left|\frac{p_0^2-(k_0-p_0)^2-\ppi^2+g'YB\Sigma_3
   -2\sqrt{\ppi^2}\sqrt{(k_0-p_0)^2-g'YB\Sigma_3}}
   {p_0^2-(k_0-p_0)^2-\ppi^2+g'YB\Sigma_3
   +2\sqrt{\ppi^2}\sqrt{(k_0-p_0)^2-g'YB\Sigma_3}}\right|\nonumber\\
   &+&\frac{\ppi\dia_\perp}{2(\ppi^2+\frac{g'Y_\phi}{2}B\Sigma_3)^{3/2}}\
   \left(p_0^2-(k_0-p_0)^2-\ppi^2-\frac{g'Y_\phi}{2}B\Sigma_3\right)\nonumber\\
   &\times&
   \ln\left|\frac{p_0^2-(k_0-p_0)^2-\ppi^2-\frac{g'Y_\phi}{2}B\Sigma_3
   -2\sqrt{\ppi^2+\frac{g'Y_\phi}{2}B\Sigma_3}\sqrt{(k_0-p_0)^2}}
   {p_0^2-(k_0-p_0)^2-\ppi^2-\frac{g'Y_\phi}{2}B\Sigma_3
   +2\sqrt{\ppi^2+\frac{g'Y_\phi}{2}B\Sigma_3}\sqrt{(k_0-p_0)^2}}\right|
   \nonumber\\
   &+&\left.\frac{2k_3b\;\dia}{\ppi^2}\sqrt{(k_0-p_0)^2-g'YB\Sigma_3}
   +\frac{2\ppi\dia_\perp}{\ppi^2+\frac{g'Y_\phi}{2}B\Sigma_3}
   \sqrt{(k_0-p_0)^2}\right\}
   \, ,\nonumber\\ 
   \hat{\Sigma}_{f\phi}&=&-\frac{1}{16\pi^2}
   \int_{-\infty}^\infty dp_0 \tilde{n}_F(p_0)\left\{
   \left[\frac{p_0u\;\dia}{\sqrt{\ppi^2}} 
   - \frac{k_3b\;\dia}{2(\ppi^2)^{3/2}}\
   \left(p_0^2-(k_0-p_0)^2+\ppi^2-g'YB\Sigma_3\right)\right]\right.\nonumber\\
   &&\ln\left|\frac{p_0^2-(k_0-p_0)^2-\ppi^2-g'YB\Sigma_3
   -2\sqrt{\ppi^2}\sqrt{(k_0-p_0)^2}}
   {p_0^2-(k_0-p_0)^2-\ppi^2-g'YB\Sigma_3
   +2\sqrt{\ppi^2}\sqrt{(k_0-p_0)^2}}\right|\nonumber\\
   &-&\frac{\ppi\dia_\perp}{2(\ppi^2+\frac{g'Y_\phi}{2}B\Sigma_3)^{3/2}}\
   \left(p_0^2-(k_0-p_0)^2+\ppi^2+\frac{g'Y_\phi}{2}B\Sigma_3\right)\nonumber\\
   &\times&
   \ln\left|\frac{p_0^2-(k_0-p_0)^2-\ppi^2-\frac{g'Y_\phi}{2}B\Sigma_3
   -2\sqrt{(k_0-p_0)^2}\sqrt{\ppi^2+\frac{g'Y_\phi}{2}B\Sigma_3}}
   {p_0^2-(k_0-p_0)^2-\ppi^2-\frac{g'Y_\phi}{2}B\Sigma_3
   +2\sqrt{(k_0-p_0)^2}\sqrt{\ppi^2+\frac{g'Y_\phi}{2}B\Sigma_3}}
   \right|\nonumber\\
   &-&\left.\frac{2k_3b\;\dia}{\ppi^2}\sqrt{(k_0-p_0)^2}
   -\frac{2\ppi\dia_\perp}{\ppi^2+\frac{g'Y_\phi}{2}B\Sigma_3}
   \sqrt{(k_0-p_0)^2}\right\}\, .
   \label{Sigmafphifphi}
\ee
\end{widetext}
As mentioned earlier, current observations suggest that the
temperature was higher as compared to the magnetic field strength
prior to the EWPT. In the next section we consider such scenario,
i.e., $T\gg\sqrt{g'B}$. 

\section{High temperature limit}\label{IV}

It is well known that the consistent picture that
systematically accounts for the leading temperature effects in
relativistic plasmas, appropriate for situations where the largest
energy scale is set by the temperature, is the so called {\it
Hard Thermal Loop} approximation~\cite{Le Bellac}. This approach
guarantees, in particular, that the temperature corrections to the
pole of particle propagators are gauge independent. The approach
remains valid also in the presence of magnetic fields~\cite{Elmfors2}
and in the present context is implemented by recalling that the
leading contribution to expressions such as Eqs.~(\ref{SigmaW2})
and~(\ref{Sigmafphifphi}) comes from large momenta (of order $T$) in
the integrand. In this approximation, terms proportional to the factor
$(1-1/\xi)$ originating from Eq.~(\ref{gaugebosprop}) are sub-leading
and can be ignored. Therefore, considering only the large $p_0$ region
in Eqs.~(\ref{SigmaW2}) and~(\ref{Sigmafphifphi}), and using that  
\be
   \int_{-\infty}^\infty dp_0 p_0 n_B(p_0)
   &=&\frac{\pi^2T^2}{3}\, ,\nonumber\\
   \int_{-\infty}^\infty dp_0 p_0 \tilde{n}_F(p_0)
   &=&\frac{\pi^2T^2}{6}\, ,
   \label{approxff}
\ee
the left- and right-handed fermion self-energies become
\begin{widetext}
\be
   \hat{\Sigma}_{L} &=&
   \frac{m_L^2}{k_0\sqrt{(\ppi^2)^L}}Q\left(k_0,\sqrt{(\ppi^2)^L}
   \right)k_0\gamma_0 + 
   \frac{m_L^2}{(\ppi^2)^L}S\left(k_0\,\sqrt{(\ppi^2)^L}\right)
   k_3\gamma_3\nonumber\\ 
   &+& \left[
   \frac{m_L^2}{(\ppi^2)^L+g'\frac{Y_L}{2}B\sigma_3}
   S\left(k_0,\sqrt{(\ppi^2)^L+g'\frac{Y_L}{2}B\sigma_3}\right)\right.
   \nonumber\\  
   &+&
   \left.\frac{g_\phi^2T^2}{16}
   \left(\frac{S\left(k_0,\sqrt{(\ppi^2)^L+g'\frac{Y_\phi}{2}
   B\sigma_3}\right)}
   {(\ppi^2)^L+g'\frac{Y_\phi}{2}B\sigma_3} - 
   \frac{S\left(k_0,\sqrt{(\ppi^2)^L+g'\frac{Y_L}{2}B\sigma_3}\right)}
   {(\ppi^2)^L+g'\frac{Y_L}{2}B\sigma_3}\right)
   \right]\left(\ppi^L\cdot\gamma\right)_\perp\, ,
   \nonumber\\
   \hat{\Sigma}_{R} &=&
   \frac{m_R^2}{k_0\sqrt{(\ppi^2)^R}}Q\left(k_0,\sqrt{(\ppi^2)^R}
   \right)k_0\gamma_0 + 
   \frac{m_R^2}{(\ppi^2)^R}S\left(k_0\,\sqrt{(\ppi^2)^R}\right)k_3
   \gamma_3\nonumber\\ 
   &+&\left[
   \frac{m_R^2}{(\ppi^2)^R+g'\frac{Y_R}{2}B\sigma_3}
   S\left(k_0,\sqrt{(\ppi^2)^R+g'\frac{Y_R}{2}B\sigma_3}\right)\right.
   \nonumber\\  
   &+&
   \left.\frac{g_\phi^2T^2}{8}
   \left(\frac{S\left(k_0,\sqrt{(\ppi^2)^R+g'\frac{Y_\phi}{2}B\sigma_3}\right)}
   {(\ppi^2)^R+g'\frac{Y_\phi}{2}B\sigma_3} - 
   \frac{S\left(k_0,\sqrt{(\ppi^2)^R+g'\frac{Y_R}{2}B\sigma_3}\right)}
   {(\ppi^2)^R+g'\frac{Y_R}{2}B\sigma_3}\right)
   \right]\left(\ppi^R\cdot\gamma\right)_\perp\, ,
   \label{largeTL}
\ee
\end{widetext}
where 
\be
   \hat{\Sigma}_L&=&\left[\left(\frac{g'Y_L}{2}\right)^2 +
   \left(\frac{g^2}{2}\right)\right]
   \left[\hat{\Sigma}_W|_{Y_L/2} + \hat{\Sigma}_f|_{Y_L/2}\right]\nonumber\\
   &+&(g_\phi^2)
   \left[\hat{\Sigma}_\phi |_{Y_L/2} + \hat{\Sigma}_{f\phi}|_{Y_L/2}\right]
   \, ,\nonumber\\
   \hat{\Sigma}_R&=&\left(\frac{g'Y_R}{2}\right)^2
   \left[\hat{\Sigma}_W|_{Y_R/2} + \hat{\Sigma}_f|_{Y_R/2}\right]\nonumber\\
   &+&2g_\phi^2\left[\hat{\Sigma}_\phi |_{Y_R/2} + 
   \hat{\Sigma}_{f\phi}|_{Y_R/2}\right]\, ,
   \label{izqder}
\ee
and where we have defined the thermal left- and right-handed masses
by~\cite{Weldon} 
\be
   m_L^2 &=& \left[\left(\frac{g'Y_L}{2}\right)^2
   + \frac{g^2}{2} + \frac{g_\phi^2}{2}
   \right]\left(\frac{T^2}{8}\right)\, ,\nonumber\\
   m_R^2 &=& \left[\left(\frac{g'Y_R}{2}\right)^2
   + g_\phi^2\right]\left(\frac{T^2}{8}\right)\, ,
   \label{thermLR}
\ee 
and the functions $Q$ and $S$ by
\be
   Q\left(k_0,\sqrt{\ppi^2}\right)&=&\frac{1}{2}
   \ln\left|\frac{k_0+\sqrt{\ppi^2}}{k_0-\sqrt{\ppi^2}}\right|\, , 
   \nonumber\\
   S\left(k_0,\sqrt{\ppi^2}\right)&=&
   1 - \frac{k_0}{2\sqrt{\ppi^2}}
   \ln\left|\frac{k_0+\sqrt{\ppi^2}}{k_0-\sqrt{\ppi^2}}\right|\, .
   \label{QandS}
\ee
Armed with the expressions for the right- and left-handed
self-energies, we are now in the position to compute the dispersion
relation. We first need to find the appropriate wave functions upon
which the above operators act. We begin next section finding such
solutions. \\

\section{Chiral fermion dispersion relations}\label{V}

We work explicitly in the chiral representation of the
$\gamma$-matrices, namely,
\be
   \gamma^0=\!\left(\begin{array}{rr}
   0 & -I \\
   -I & 0 \end{array}\right)\
   \mbox{\boldmath $\gamma$}=\!\left(\begin{array}{rr}
   0 &  \mbox{\boldmath $\sigma$} \\
   \mbox{\boldmath $-\sigma$} & 0 \end{array}\right)\
   \gamma_5=\!\left(\begin{array}{rr}
   I & 0 \\
   0 & -I \end{array}\right)\, .
   \label{gammaschiral}
\ee
In order to find the dispersion relation for left- and right-handed
modes, let us look for the eigenvalues of the effective Dirac equation
\be
   (\ppi\dia - \hat{\Sigma} )\Psi = 0\, ,
   \label{diraceff}
\ee
where $\hat{\Sigma}=\hat{\Sigma}_L\ L + \hat{\Sigma}_R\ R$. A suitable
ansatz for the solution to Eq.~(\ref{diraceff}) is found by noticing
that inclusion of self-interactions does not alter the chiral nature
of the equation. Thus, the solution is found in the same manner as for
the case when $\hat{\Sigma}=0$ and is given in cylindrical coordinates
by~\cite{Ayala2} 
\begin{widetext}
\be
   \Psi(r,\phi, z)\equiv\left(
                    \begin{array}{c}
                    \Psi_1^R\\ \Psi_2^R\\ \Psi_1^L\\ \Psi_2^L    
                    \end{array}\right)=
                    \left(
                    \begin{array}{c}
                       \left\{
                       \begin{array}{c}
                          C_1^RI_{n-1,s}(\rho^R)e^{i(l-1)\phi}\\
                          iC_2^RI_{n,s}(\rho^R)e^{il\phi}
                       \end{array}\right\}e^{ik^Rz}\\
                       \left\{
                       \begin{array}{c}
                          C_1^LI_{n-1,s}(\rho^L)e^{i(l-1)\phi}\\
                          iC_2^LI_{n,s}(\rho^L)e^{il\phi}
                       \end{array}\right\}e^{ik^Lz}
                    \end{array}\right)\, ,
   \label{finalpsisym}
\ee
\end{widetext}
where $C_1^{R,L}$ and $C_2^{R,L}$ are constants,
\be
   \rho^{R,L}&\equiv&\gamma^{R,L}r^2\nonumber\\
   \gamma^{R,L}&\equiv& \frac{g'Y^{R,L}}{4}B\, ,
   \label{defs}
\ee
and $I_{n,s}(\rho)$ are the
Laguerre functions given in terms of the Laguerre polynomials
$Q_s^{n-s}$ by 
\be
   I_{n,s}(\rho)=\frac{1}{(n!s!)^{1/2}}e^{-\rho/2}\rho^{(n-s)/2}
   Q_s^{n-s}(\rho)\, ,
   \label{Laguerre}
\ee
satisfying the normalization condition
\be
   \int_0^\infty d\rho I_{n,s}^2(\rho)=1\, .
   \label{norm}
\ee
$n=l+s$ is called the principal quantum number that labels the
corresponding Landau level and must be a non-negative
integer. Defining
\be
   -i\ppi_\pm^{R,L} = \ppi_1^{R,L}\pm i\ppi_2^{R,L}\, ,
   \nonumber
   \label{pipm}
\ee 
it is easy to show that
\be
   \ppi_+^{R,L} I_{n-1,s}&=&-\sqrt{4n\gamma^{R,L}}I_{n,s}\nonumber\\
   \ppi_-^{R,L} I_{n,s}&=&\sqrt{4n\gamma^{R,L}}I_{n-1,s}\nonumber\\
   (\ppi^2_\perp )^{R,L} &=& \ppi_-^{R,L}\ppi_+^{R,L} - 
   \frac{g'Y^{R,L}}{2}B\, ,
   \label{pipmprpo}
\ee
and therefore 
\begin{widetext}
\be
   \left((\ppi_\perp^2)^{R,L} + \frac{g'Y^{R,L}}{2}B\sigma_3\right)
   \left\{\begin{array}{c}
                          \Psi_1^{R,L}\\
                          \Psi_2^{R,L}
                       \end{array}\right\}&=&
   (4\gamma^{R,L} n)
   \left\{\begin{array}{c}
                          \Psi_1^{R,L}\\
                          \Psi_2^{R,L}
                       \end{array}\right\}\, ,\nonumber\\
   \left((\ppi_\perp^2)^{R,L} + \frac{g'Y^{R,L}}{2}B\sigma_3\right)
   \left\{\begin{array}{c}
                          \Psi_1^{R,L}\\
                          \Psi_2^{R,L}
                       \end{array}\right\}&=& 
   \left\{\begin{array}{c}
   g'\left[(n-\frac{1}{2})Y^{R,L} + \frac{Y_\phi}{2}\right]B \Psi_1^{R,L}\\
   g'\left[(n-\frac{1}{2})Y^{R,L} + \frac{Y_\phi}{2}\right]B \Psi_2^{R,L}
                       \end{array}\right\}\, .
   \label{piperpoverI}
\ee
\end{widetext}
Using Eqs.~(\ref{thermLR}) and~(\ref{piperpoverI}) into the effective
Dirac equation, Eq.~(\ref{diraceff}), we find the conditions for
the existence of self-consistent solutions in the form of the secular
equations 
\be
   \det \Delta^{R,L}=0\,,
   \label{det}
\ee
where $\Delta^{R,L}$ are given by
\begin{widetext}
\be
   \Delta^{R,L}=
   \left(
      \begin{array}{cc}
          U^{R,L}_+\left(\sqrt{g'(n-\frac{1}{2})Y^{R,L}B + k_3^2}\ \right)&
          V^{R,L}\left(
          \sqrt{g'\left[(n-\frac{1}{2})Y^{R,L} + \frac{Y_\phi}{2}\right]B 
          + k_3^2}\ \right)\\
          V^{R,L}\left(
          \sqrt{g'\left[(n+\frac{1}{2})Y^{R,L} + \frac{Y_\phi}{2}\right]B 
          + k_3^2}\ \right)&
          U^{R,L}_-\left(\sqrt{g'(n+\frac{1}{2})Y^{R,L}B + k_3^2}\ \right)
      \end{array}
   \right)\, ,
   \label{deltas}
\ee
\end{widetext}
and the functions $U_\pm$ and $V_\pm$ are given by
\begin{widetext}
\be
   U^{R,L}_\pm (x)&=&k_0\left[1-\frac{m_{R,L}^2}{k_0x}
   Q\left(k_0,x\right)\right]
   \pm k_3\left[1+\frac{m_{R,L}^2}{x^2}
   S\left(k_0,x\right)\right]\nonumber\\
   V^L(x)&=&\sqrt{g'nY^LB}\
   \left\{1 + \frac{(m_L^2 - g_\phi^2T^2/16)}{(g'nY^LB + k_3^2)}\
   S\left(k_0,\sqrt{g'nY^LB + k_3^2}\right)
   + \left(\frac{g_\phi^2T^2}{16}\right)
   \frac{S\left(k_0,x\right)}{x^2}
   \right\}\nonumber\\
   V^R(x)&=&\sqrt{g'nY^RB}\
   \left\{1 + \frac{(m_R^2 - g_\phi^2T^2/8)}{(g'nY^RB + k_3^2)}\
   S\left(k_0,\sqrt{g'nY^RB + k_3^2}\right)
   + \left(\frac{g_\phi^2T^2}{8}\right)
   \frac{S\left(k_0,x\right)}{x^2}
   \right\}\, .
   \label{UandV}
\ee
\end{widetext}
Equation~(\ref{det}) can be solved to find the dispersion relation
$k_0=\omega (k_3)$, for any given Landau level. It is
convenient however to pause for a moment and discuss the 
dispersion relation for modes occupying the lowest Landau level (LLL),
$n=0$. In this case $I_{n-1,s}\rightarrow I_{-1,s}$ and since the
Laguerre functions vanish when the principal quantum number is a
negative integer, the secular equation, Eq.~(\ref{det}), becomes
simply
\be
   U^{R,L}_-\left(\sqrt{\frac{g'}{2}Y^{R,L}B + k_3^2}\ \right)=0\, .
   \label{secn0}
\ee
Equation~(\ref{secn0}) gives the dispersion relation for left- and
right-handed modes in the LLL. Recall that the high-temperature
dispersion relation for fermions consists of two
branches~\cite{Le Bellac} usually referred to as the {\it particle} and
the {\it hole} solutions. The hole is in fact a positive energy
solution that in the absence of thermal effects would just correspond
to a negative energy solution. 

\begin{figure}[t!] 
\vspace{0.5cm}
{\centering
\resizebox*{0.48\textwidth}{0.24\textheight}
{\includegraphics{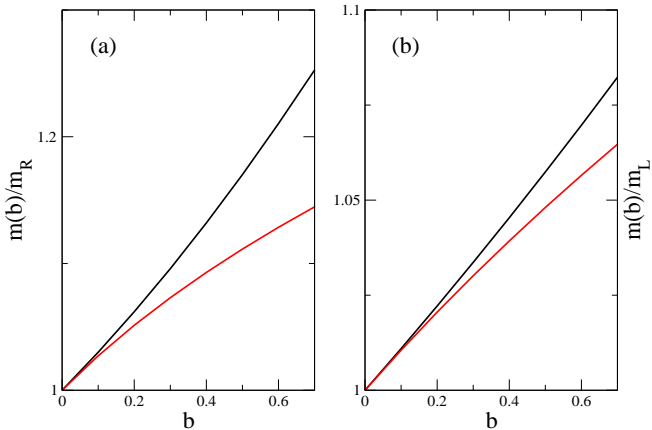}}
\par}
\caption{Magnetic field dependence of the particle's mass in the
lowest Landau level as a function of $B$ parametrized as $B=bT^2$ for
(a) right-handed and (b) left-handed modes. Upper curves are the
exact solutions and lower curves the analytical approximations.}
\label{fig2}
\end{figure}
\vspace{1cm}

The fact that for each mode there is only one non-vanishing solution
to Eq.~(\ref{secn0}) means that only one
projection of the spin relative to the direction of the external
fields is able to propagate. This is precisely what
one should expect for modes occupying the lowest possible energy
level. The potential energy of a particle interacting with the external
field is
\be
   U=-\vec{\mu}\cdot\vec{B}\propto
   -Y^{R,L}\vec{S}\cdot\vec{B}\, , 
   \label{potenergy}
\ee
with $\vec{\mu}$, the particle's magnetic moment and $\vec{S}$ the
particle's spin. Suppose the mode refers to a particle solution
for a top-quark, then in Eq.~(\ref{potenergy}),
$Y^{R,L}>0$ . Thus, $U$ becomes a  minimum for $\vec{S}$ and 
$\vec{B}$ parallel. Since the modes are 
chiral and also eigenfunctions of the helicity operator, left-
(right)-handed particles can satisfy this condition only if $k_3<0$
$(k_3>0)$. This means that in the LLL, left-
(right)-handed particles can only exist moving in the direction
opposite (parallel) to the direction of the external field, with modes
with the opposite value of $k_3$ being forbidden. For hole
solutions the converse applies.

The dependence of the thermal mass on the magnetic field strength for
the LLL modes can also be computed by considering the limit when
$k_3\rightarrow 0$ in Eq.~(\ref{secn0}) which then becomes explicitly
\be
   k_0 - \frac{m^2_{R,L}}{\sqrt{\frac{g'}{2}Y^{R,L}B}}\
   \frac{1}{2}\ln\left|\frac{k_0+\sqrt{\frac{g'}{2}Y^{R,L}B}}
   {k_0-\sqrt{\frac{g'}{2}Y^{R,L}B}}\right| = 0\, .
   \label{kz0}
\ee
In order to obtain analytical results, let us consider the limit in
which $k_0\sim g'T\gg \sqrt{g'B}$. Then, we can use the Taylor
expansion
\be
   \frac{1}{2}\ln\left(\frac{1+x}{1-x}\right)\simeq x+\frac{x^3}{3}\, ,
   \label{expan}
\ee
into Eq.~(\ref{kz0}) and obtain
\be
   m_{R,L}(B)&\equiv&k_0^{R,L}(k_3=0)\nonumber\\
   &=&\frac{m_{R,L}}{\sqrt{2}}
   \left(1+\sqrt{1+\frac{2g'Y^{R,L}B/3}{m^2_{R,L}}}\ \right)^{1/2}
   \!\!\!\!\!\!\!\!.
   \label{koaprox}
\ee
We thus see that the effect of the magnetic field is to increase the
value of the thermal mass of the modes occupying the LLL. The
increase however is different for left- and right-handed fields as the
hypercharge for these modes is different. We also notice that in the
limit when $B\rightarrow 0$, $m(B)_{R,L}\rightarrow m_{R,L}$. 

Figure~\ref{fig2} shows the behavior of the solutions of
Eq.~(\ref{kz0}) as functions of $B$ parametrized by $B=bT^2$. We have
taken $g'=0.344$, $g=0.637$ as corresponds to the values of these
coupling constants at the EWPT epoch, as well as $g_\phi=1$,
$Y_R=\frac{\mbox\small {4}}{\mbox\small {3}}$ and
$Y_L=\frac{\mbox\small {1}}{\mbox\small {3}}$, as appropriate for top
quarks. Also shown in the figure is the approximate analytical solution,
Eq.~(\ref{koaprox}). Notice that the difference between the exact and
the approximate results grows with $b$ as is to be expected since the
approximate solution, Eq.~(\ref{koaprox}), is only valid for values of
$b$ satisfying $g'\gg b$.

\begin{figure}[t!] 
\vspace{0.5cm}
{\centering
\resizebox*{0.48\textwidth}{0.24\textheight}
{\includegraphics{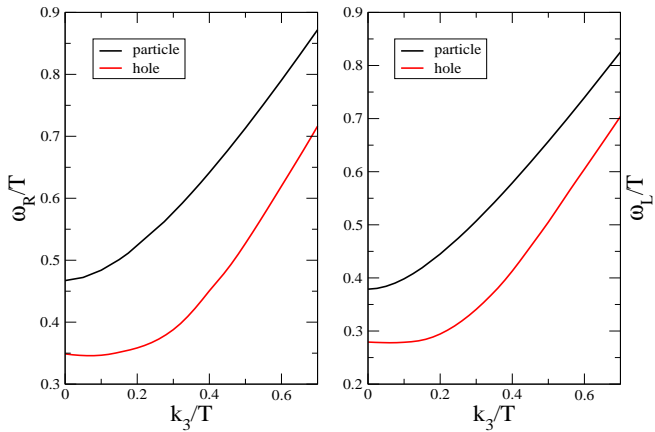}}
\par}
\caption{Dispersion relation as a function of the longitudinal
momentum for right-handed (left panel) and left-handed (right panel)
modes for $n=1$ and $b=0.1$. Upper curves are the particle solutions
and lower curves the hole solutions. Notice that the presence of the
magnetic field breaks the degeneracy for the mass of particle and hole
solutions, pushing the former to larger and the latter to lower
values, as compared to the thermal mass in the absence of a magnetic
field, for the given value of $b$.}
\label{fig3}
\end{figure}
\vspace{1cm}

Figure~\ref{fig3} shows the dispersion relation for right- and
left-handed modes in the Landau level with $n=1$. We have used a value
for the magnetic field with $b=0.1$. Notice that the presence of the
magnetic fields breaks the degeneracy in mass for particle and hole
solutions. For the value of $b$ considered, the effect of the magnetic
field is to increase the mass for particle and reduce it for hole
solutions as compared to the value of the corresponding thermal mass in
the absence of a magnetic field. The mass splitting is different
however for right- and left-handed modes due to their different
couplings to the external field. 

\begin{figure}[t!] 
\vspace{0.5cm}
{\centering
\resizebox*{0.48\textwidth}{0.24\textheight}
{\includegraphics{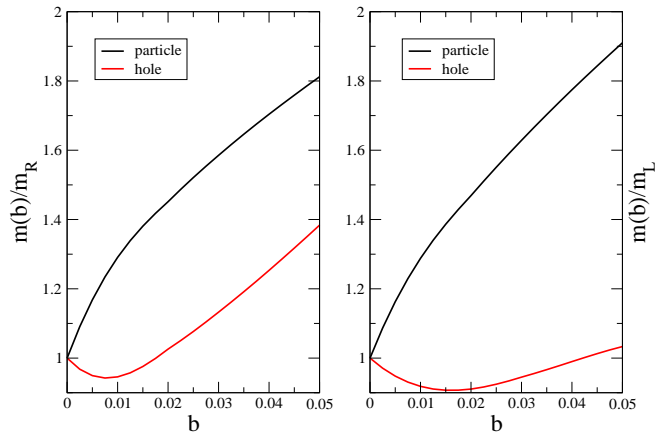}}
\par}
\caption{Magnetic field dependence of the particle's mass in the
n=10 Landau level as a function of $B$ parametrized as $B=bT^2$ for
right-handed (left panel) and left-handed (right panel) modes. Upper
curves are the particle and lower curves the hole solutions.}
\label{fig4}
\end{figure}

Figure~\ref{fig4} shows the magnetic field dependence of the
particle's mass in the Landau level with n=10 as a function of $B$
parametrized as $B=bT^2$ for right- and left-handed modes. The mass
splitting between particle and hole solutions is more important for
left-handed modes for which the value of the hypercharge is smaller
than the corresponding value for right-handed modes.

\section{Summary and conclusions}\label{concl}

In this work we have computed the dispersion relation for chiral
fermions in the symmetric phase of the electroweak theory in the
presence of a constant hypermagnetic field, at one loop level but all
orders in $g'B$. Working in the limit $T\gg \sqrt{g'B}$, we have shown
that left- and right-handed modes occupying the same Landau level,
develop finite but different thermal masses due to the chiral nature
of their coupling to the external field. In particular, in the LLL the
thermal mass of the modes is increased with increasing field
strength. For the rest of the levels with $n\neq 0$, the hypermagnetic
field breaks the mass degeneracy for particle and hole
solutions, however the mass splitting is different for left- and
right-handed modes as their couplings to the external fields are
different. 

We have argued that, given the current bounds on the
strength of primordial magnetic fields, the large temperature, weak
field limit corresponds to the relevant scenario for the propagation
of fermions prior and during the electroweak phase transition.

\vspace{1cm}

Though the numerical calculations are performed for top quarks, the
results can also be applied to the case of any SM fermion species that
can couple minimally to hypermagnetic fields prior to the EWPT through
a non-vanishing hypercharge. Our work shows that the motion of such
fermions is highly anisotropic since for any given Landau level, it is
directed along the field lines. Neutrinos can prove to be another 
interesting case since at their decoupling, this anisotropic motion should
be reflected in the properties of the cosmic background of relic
neutrinos. Thus, if these relic neutrinos were to be detected and the
anisotropy measured, this would provide  a means of confirming the
existence of primordial magnetic fields in the early universe.
Another interesting consequence is the asymmetry
that can be generated in decay processes of chiral fermions where
particles with only one chirality are produced, such as beta decay. This
is best illustrated for the case of the lowest Landau level where only
one direction of motion is allowed for a given chirality of the decay
products. The same asymmetry is present in higher Landau levels. 

\section*{Acknowledgments}

A.A. wishes to thank L. McLerran for his kind hospitality during a
summer visit to BNL where part of this work was completed. The authors
are indebted to V. de la Incera and E. Ferrer for very useful
comments. Support for this work has been received in part by PAPIIT
under grants number IN108001 and IN109001 and CONACyT under grants
number 32395-E, 35792-E and 40025-F.


\begin{thebibliography}{55}

\bibitem{Reviews}
For reviews on the origin, evolution and some cosmological
consequences of primordial magnetic fields see: K. Enqvist,
Int. J. Mod. Phys. {\bf D7}, 331 (1998); R. Maartens, {\it Cosmological
magnetic fields}, {\it International Conference on Gravitation
and Cosmology}, Pramana {\bf 55}, 575 (2000) and references therein;  
D. Grasso and H.R. Rubinstein, Phys. Rep. {\bf 348}, 163 (2001);
L.M. Widrow, Rev. Mod. Phys. {\bf 74}, 775 (2003). 
 
\bibitem{Kron}
See for example P.P. Kronberg, Rep. Prog. Phys. {\bf 57}, 325
(1994); J.-L. Han and R. Wielebinski, {\it Milestones in the
Observations of Cosmic Magnetic Fields}, astro-ph/0209090. 

\bibitem{Beck}
R. Beck, A. Brandenburg, D. Moss, A. Shukurov and D. Sokoloff,
Annu. Rev. Astron. Astrophys. {\bf 34}, 155 (1996).

\bibitem{Eilek}
J. A. Eilek and F. N. Owen, Ap. J. {\bf 567}, 202 (2002).

\bibitem{Clarke}
T. E. Clarke, P. P. Kronberg and H. B\"ohringer, Ap. J. {\bf 547},
L111 (2001). 

\bibitem{Giovannini1}
M. Giovannini, {\it Pri\-mor\-dial Mag\-netic Fields}, hep-ph/0208152.

\bibitem{Quash}
J. Quashnock, A. Loeb and D.N. Spergel, Ap. J. {\bf 344}, L49 (1989);
B. Cheng and A.V. Olinto, Phys. Rev. D {\bf 50}, 2421 (1994); G. Sigl,
A.V. Olinto and K. Jedamzik, Phys. Rev. D {\bf 55}, 4582 (1997). 

\bibitem{Baym}
G. Baym, D. B\"odeker and L. McLerran, Phys. Rev. D
{\bf 53}, 662 (1996).  

\bibitem{Boyan}
D. Boyanovsky, H. J. de Vega and M. Simionato, {\it Large scale
magnetogenesis from a non-equilibrium phase transition in the
radiation dominated era}, hep-ph/0211022. 

\bibitem{Ayala1} 
A. Ayala, J. Besprosvany, G. Pallares and G. Piccinelli, \prd {\bf
64}, 123529 (2001); A. Ayala, G. Piccinelli and G. Pallares,  \prd {\bf
66}, 103503 (2002). 

\bibitem{Ayala2}
A. Ayala and J. Besprosvany, Nucl. Phys. B {\bf 651}, 211 (2003).

\bibitem{Cannellos}
J. Cannellos, E.J. Ferrer, V. de la Incera, Phys. Lett. B {\bf 542},
123 (2002), E.J. Ferrer and V. de la Incera, {\it Neutrinos under
strong magnetic fields}, hep-ph/0308017.

\bibitem{Barrow} J.D. Barrow, P.G. Ferreira and J. Silk,
Phys. Rev. Lett. {\bf 78}, 3610 (1997).

\bibitem{Schwinger}
J. Schwinger, Phys. Rev. {\bf 82}. 664 (1951).

\bibitem{Sahu}
J.C. D'Olivo, J.F. Nieves and S. Sahu, Phys. Rev. D {\bf 67}, 025018
(2003). 

\bibitem{Dittrich}
W. Dittrich and M. Reuter, {\it Effective Lagrangians in Quantum
Electrodynamics}, Lecture Notes in Physics (Springer-Verlag, Berlin, 1985).

\bibitem{Le Bellac}
See for example, M. Le Bellac, {\it Thermal Field Theory} (Cambridge
University Press, Cambridge, 1989).

\bibitem{Elmfors2}
P. Elmfors, D. Persson and B.-S. Skagerstam, Nucl. Phys. B {\bf 464},
153 (1996)

\bibitem{Weldon}
Similar expressions were first obtained by H.A. Weldon, Phys. Rev. D
{\bf 26}, 2789 (1982). 

\end{thebibliography}
\end{document}